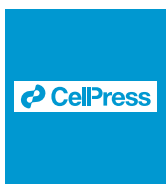
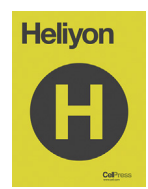

Research article

# Assessment of global warming in Al Buraimi, sultanate of Oman based on statistical analysis of NASA POWER data over 39 years, and testing the reliability of NASA POWER against meteorological measurements

Osama A. Marzouk[*]

*College of Engineering, University of Buraimi, P.O. Box 890, Post Code 512, Al Buraimi, Sultanate of Oman*

ABSTRACT

We performed a number of statistical analysis methods on the historical data for the air temperature at 2 m above the ground and its range, as reported by the database of NASA known as POWER, which stands for Prediction Of Worldwide Energy Resources. The point of analysis is the University of Buraimi, located in Al Buraimi Governate, in the Northwest of the Sultanate of Oman, near its border with United Arab Emirates (UAE). The data is in the form of a value per day, for every day in the year. The data analyzed span the period from January 3$^{rd}$, 1981 (earliest day available) to December 31$^{st}$, 2019 (latest end-of-year available).

The statistical analysis methods include: simple linear regression, F-test: two-sample for variances, analysis of variance (ANOVA): single factor, and t-test: two-sample assuming equal variances (pooled).

The results show that the mean of the local 2-meter air temperature is increasing at a rate of about 0.039 °C per year, starting from an estimated value of 27.4 °C in 1980.

For the standard deviation of the 2-meter air temperature, and the mean and standard deviation of its range; although a linear regression analysis suggests a decline over time, the regression coefficient is not significant. On the other hand, the analysis of variance for the 9 years 1981, 1985, 1990, 1995, 2000, 2005, 2010, 2015, 2019 suggests existence of statistical difference among them collectively, for the 2-meter air temperature and its range.

In the second part of this work, NASA POWER data for the air temperature at 2 m above the ground, the atmospheric pressure, the relative humidity at 2 m above the ground, and the daily precipitation were compared with recorded sensor measurements at Manah meteorological station, located in Al Dakhiliyah Governate, in the Northeast of the Sultanate of Oman over all days of years 2011, 2012, 2015, and 2016. The statistical analysis and visual inspection suggest that NASA POWER data are reliable for the 2-meter air temperature, while showing about 2.1 kPa constant shift (underestimation) for the atmospheric pressure. The data show mild inaccuracy for the 2-meter relative humidity, but are largely unreliable for the precipitation, with significantly exaggerated values compared to real recordings.

## 1. Introduction

Climate change due to anthropogenic (human-caused) activities is a global challenge for humans, both as governments and as individuals (Butler, 2018). While climate change refers to a long-term deviation in normal weather patterns locally and globally due to both anthropogenic effects and natural effects, the term global warming focuses on the observed gradual temperature increase of Earth's climate since the pre-industrial period (before 1750) as a result of anthropogenic effects, mainly the combustion of fossil fuels. These human activities increase the level of greenhouse gases in the Earth's atmosphere, leading to less chance of incoming solar heat to escape back to the space when re-radiated from the Earth (at a lower temperature and frequency). Despite this small difference in the meaning, it is common to use "climate change" and "global warming" interchangeably. Usually, global warming is quantified through the average increase in the global surface temperature of the Earth (NASA, 2020a).

Several studies or published data (Karl et al., 2015; Letcher, 2019; IPCC, 2020; Colombo et al., 2020) have disproved the claim of "global warming hiatus" (Medhaug et al., 2017) in reference to a slowdown or a pause in the increase of global temperatures, or have treated the global warming as an occurring fact.

* Corresponding author.
 *E-mail address:* osama.m@uob.edu.om.

There has been a global annual temperature increase with an average rate of 0.007 °C since 1880, which grew to 0.017 °C since 1970 (NOAA, 2016). The average global surface temperature for land and ocean for the period January–July 2020 was 14.85 °C, which is 1.05 °C above the 20th century average of 13.8 °C. The 2020 January–July period was the second highest January–July period on record, with the highest being 14.89 °C in 2016 (differing with marginal gap of 0.04 °C). In chronological order, the warmest 10 years on record covering 1880–2019 are:

- (tied as 10th) 1998 and 2009
- (9th) 2005
- (7th) 2010
- (8th) 2013
- (6th) 2014
- (3rd) 2015
- (1st) 2016
- (4th) 2017
- (5th) 2018
- (2nd) 2019

It is noted that all completed years starting 2013 are in the list (NOAA, 2020).

While climate change is not a new phenomenon, the scientific study and attempts to explain it goes back about two centuries. Chao and Feng (2018) provide an overview about the historical development of the modern climate change science, which started in 1824 when Jean-Baptiste Joseph Fourier (1824) suggested that the Earth's atmosphere serves like an envelope of a greenhouse to increase the temperature. Fourier is the French mathematician and physicist known for the Fourier series in mathematics and Fourier's law in heat transfer (Lienhard IV and Lienhard V, 2019). The observation of carbon dioxide concentration by the Mauna Loa Observatory in Hawaii, USA was a milestone in the progress of research on climate change.

Informal exchange of views about global warming within the circle of family members and friends was found to encourage individuals to gain knowledge about influential facts, such as the trueness of anthropogenic global warming. In turn, better acceptance of scientific consensus of global warming increases attention toward climate change (Goldberg et al., 2019).

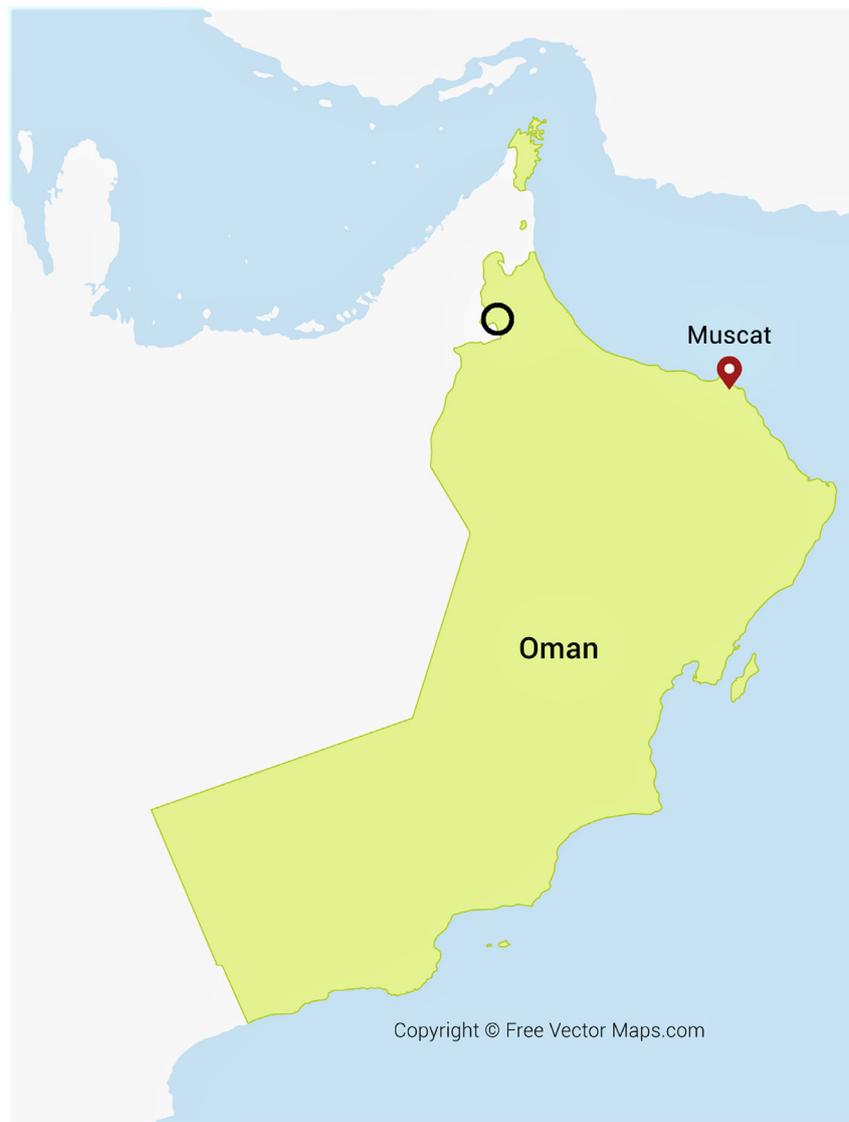

**Figure 1.** Map of the Sultanate of Oman (or simply "Oman"). Original map by FreeVectorMaps.com FreeVectorMaps.com. The original map is adapted here through (1) cropping, (2) shifting the copyright notice location, (3) enhancing the text clarity and color contrast, (4) adding a hollow circle in the Northwest of Oman to show the location of the University of Buraimi within Al Buraimi (or simply "Buraimi") Omani Governate.

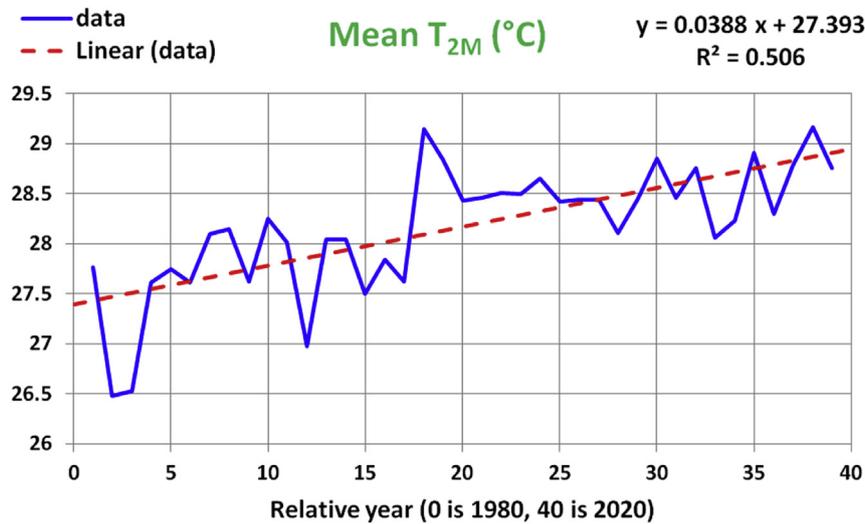

**Figure 2.** Historical data and linear regression line for the annual-mean values of the 2-meter air temperature. The x-axis represents years from 1980 to 2020.

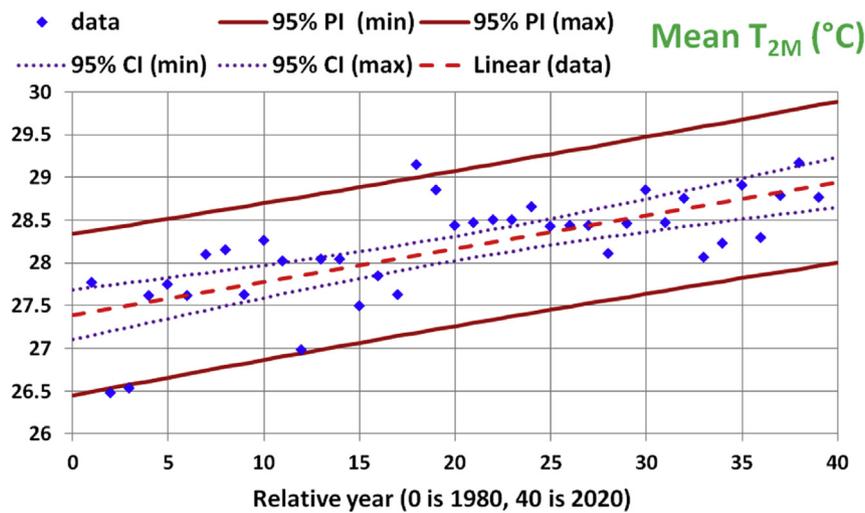

**Figure 3.** Historical data and linear regression line for the annual-mean values of the 2-meter air temperature. The 95% prediction interval (PI) bounds are highlighted with solid lines. The 95% confidence interval (CI) bounds are highlighted with dotted lines. The x-axis represents years from 1980 to 2020.

**Table 1.** ANOVA test for the linear regression model for the mean of $T_{2M}$.

|  | df | SS | MS | F | p-value |
|---|---|---|---|---|---|
| Regression | 1 | 7.4234633 | 7.42346 | 37.89 | 3.86e-7 |
| Residual | 37 | 7.2482563 | 0.1958988 |  |  |
| Total | 38 | 14.6717196 |  |  |  |

Despite the negative impacts of climate change, such as elevated global drought intensity (Vicente-Serrano et al., 2020), increased global economic inequality (Diffenbaugh and Burke, 2019), expansion of drylands (Koutrouli, 2019), and higher frequency of precipitation extremes (Papalexiou and Montanari, 2019), it serves as an alarm to individuals and institutions, encouraging rapid transition to environment-friendly alternatives in performing human activities, such as more adoption of clean energy and less utilization of fossil fuels (Acar and Dincer, 2020).

This work is concerned about the local community of Al Buraimi (or simply "Buraimi"), an inland Governate (the major political division in the country) in the Sultanate of Oman (or Oman) in terms of the global warming evidence and (if confirmed) its rate. Figure 1 is a map for Oman with an indication of Buraimi. It is the Northwest Governate of the 10 contiguous Governates of Oman. The 11th Governate is Musandam in the North, detached from the main part of Oman (Oman Media Portal, 2020). The area of Buraimi is 7,460 km². The estimated population was 45,688 in July 1, 2011; 111,394 in July 1, 2016; and 115,658 in July 1, 2019 (City Population, 2020). It consists of 3 districts (the local term is pronounced as "wilayat" or "wilayah") as: Al Buraimi (the largest one), Mahdhah, and As-Sunaynah.

**Table 2.** t-test of significance for the linear regression model for the mean of $T_{2M}$.

|  | Coefficients | Standard Error | t Stat | p-value | Lower 95% | Upper 95% |
|---|---|---|---|---|---|---|
| $b_0$ | 27.393 | 0.144517 | 189.5497 | 6.96e-57 | 27.1004 | 27.6861 |
| $b_1$ | 0.038765 | 0.00629727 | 6.155841 | 3.86e-7 | 0.026006 | 0.051524 |

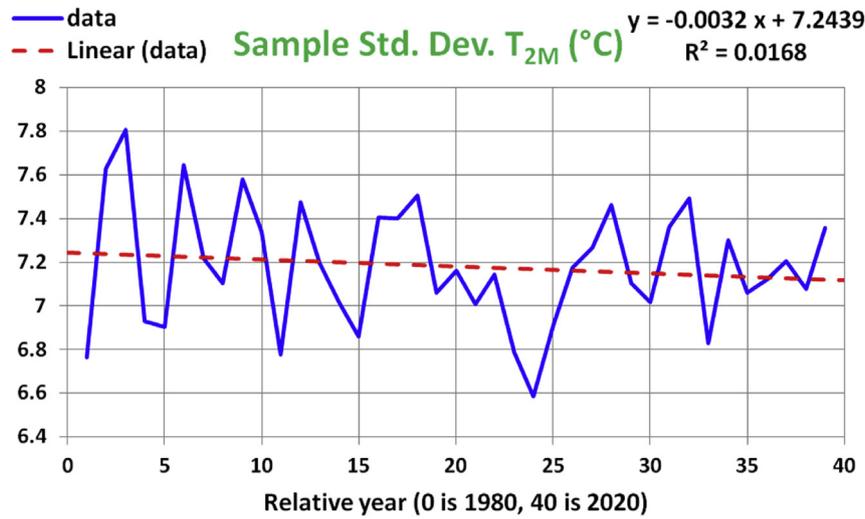

**Figure 4.** Historical data and linear regression line for the annual sample-standard-deviation values of the 2-meter air temperature. The x-axis represents years from 1980 to 2020.

## 2. Methodology

The raw data used for assessing the local warming are those taken from the online public database called POWER by NASA (NASA, 2020b). The term POWER is an acronym for Prediction Of Worldwide Energy Resources. It is based on satellite observations and models, and it is described as sufficiently accurate for reliable solar and meteorological use, especially when surface measurements are not available or are scarce. The database has the advantage of being generally contiguous in time, and of being provided on a global grid having a resolution of 0.5° latitude by 0.5° longitude.

The specific point of analysis entered in the POWER interface was within the Building C of the University of Buraimi, with the decimal degrees coordinates of 24.233935° N (latitude) and 55.892071° E (longitude). These correspond to N24 14 02, E55 53 31 in the degrees, minutes, seconds format; N 24 14.036 E 55 53.524 in the GPS format; and 40N, 387511, 2680573 in the Universal Transverse Mercator (UTM) format (GPS Coordinate Converter, 2020).

The variables considered are the air temperature at a height of 2 m from the ground ($T_{2M}$) and its range (difference between the maximum and the minimum). The data accessed are in the form of a numerical value per day for each variable, starting from January 3rd, 1981 (earliest date possible) to December 31st, 2019 (last end-of-the-year day when this research paper was initially prepared). Thus, the data span 39 years excluding 2 days, with a total of 14,242 days. The lack of 2 days in year 1981 is ignored in the analysis, and this year (1981) with 363 data points instead of 365 data points is still considered a full year with regard to calculation of a year average or a year standard deviation, as a slight approximation. The land surface temperature (LST) may be used to assess climate change (Mustafa et al., 2020). LST varies also with factors such as elevation and solar radiation (He et al., 2019). The air temperature was utilized here, rather than LST. The NASA POWER data has an advantage of simplicity and do not require a background in GIS (geographic information system) software or processing of satellite images to extract LST.

**Table 3.** ANOVA test for the linear regression model for the sample standard deviation of $T_{2M}$.

|  | df | SS | MS | F | p-value |
|---|---|---|---|---|---|
| Regression | 1 | 0.05009865 | 0.0500986 | 0.63328 | 0.43123 |
| Residual | 37 | 2.92704739 | 0.0791094 |  |  |
| Total | 38 | 2.977146034 |  |  |  |

**Table 4.** t-test of significance for the linear regression model for the sample standard deviation of $T_{2M}$.

|  | Coefficients | Standard Error | t Stat | p-value | Lower 95% | Upper 95% |
|---|---|---|---|---|---|---|
| $b_0$ | 7.24391 | 0.0918372 | 78.8778 | 7.815e-43 | 7.05783 | 7.42999 |
| $b_1$ | -0.003185 | 0.0040018 | -0.795791 | 0.43123 | -0.011293 | 0.0049238 |

**Table 5.** ANOVA analysis for $T_{2M}$ in 9 groups represented by years 1981, 1985, 1990, 1995, 2000, 2005, 2010, 2015, and 2019.

| Source of Variation | SS | df | MS | F | p-value | Critical F |
|---|---|---|---|---|---|---|
| Between Groups | 785.1551 | 8 | 98.1444 | 1.9778 | 0.04535 | 1.9412 |
| Within Groups | 162512.1 | 3275 | 49.6220 |  |  |  |
| Total | 163297.3 | 3283 |  |  |  |  |

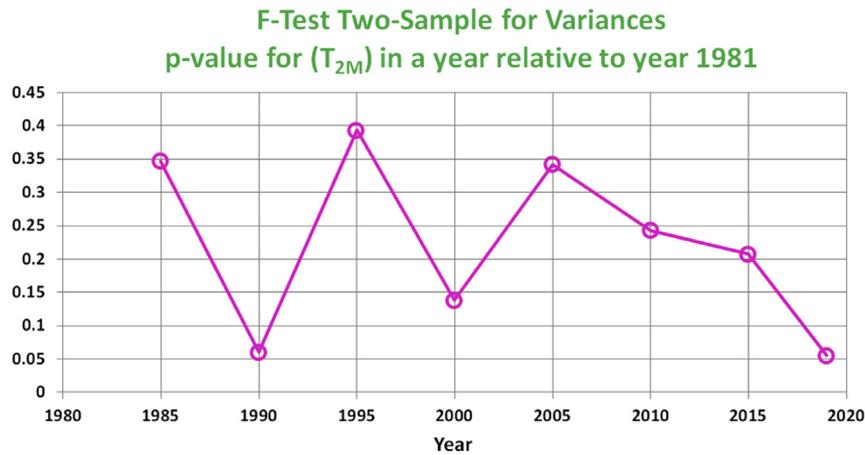

**Figure 5.** p-values of the F-test two-sample for variances of $T_{2M}$, comparing the sample variance of each year of: 1985, 1990, 1995, 2000, 2005, 2010, 2015, and 2019 with the sample variance of year 1981.

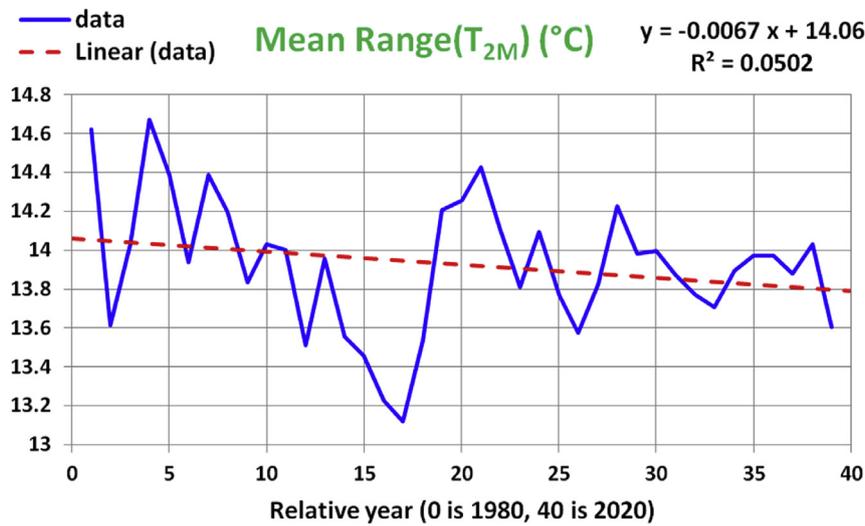

**Figure 6.** Historical data and linear regression line for the annual-mean values of the range of the 2-meter air temperature. The x-axis represents years from 1980 to 2020.

The statistical analysis (including plotting) is performed using built-in functionality or the add-in Data Analysis tools in the spreadsheet program Microsoft Excel, version 14, Microsoft Office Home and Student. The significance level ($\alpha$) is 0.05 for all relevant analyses performed here.

### 3. Results for local climate warming

The series of analysis steps starts with a scatter plot for the annual-mean of the air temperature at 2-meter height ($T_{2M}$) in Figure 2. The x-axis represents years relative to 1980. For each year (each value on the x-axis), the y-axis is the arithmetic mean of the $T_{2M}$ values ($\overline{T_{2M}}$) over all days in that year. Instead of using isolated marker points, the data values in the plot are connected with straight solid lines. The dashed line is the linear regression trend, whose formula is displayed in the figure, which translates to:

$$\overline{T_{2M}}(^\circ C) = 0.0388(\text{year} - 1980) + 27.393 \quad (1)$$

**Table 6.** ANOVA test for the linear regression model for the mean of range ($T_{2M}$).

|  | df | SS | MS | F | p-value |
|---|---|---|---|---|---|
| Regression | 1 | 0.2239485 | 0.2239485 | 1.956757 | 0.17019 |
| Residual | 37 | 4.2346288 | 0.1144494 |  |  |
| Total | 38 | 4.4585773 |  |  |  |

**Table 7.** t-test of significance for the linear regression model for the mean of range ($T_{2M}$).

|  | Coefficients | Standard Error | t Stat | p-value | Lower 95% | Upper 95% |
|---|---|---|---|---|---|---|
| $b_0$ | 14.0597 | 0.110462 | 127.281 | 1.708e-50 | 13.8359 | 14.2835 |
| $b_1$ | -0.006733 | 0.0048133 | -1.3988 | 0.17019 | -0.016486 | 0.0030197 |

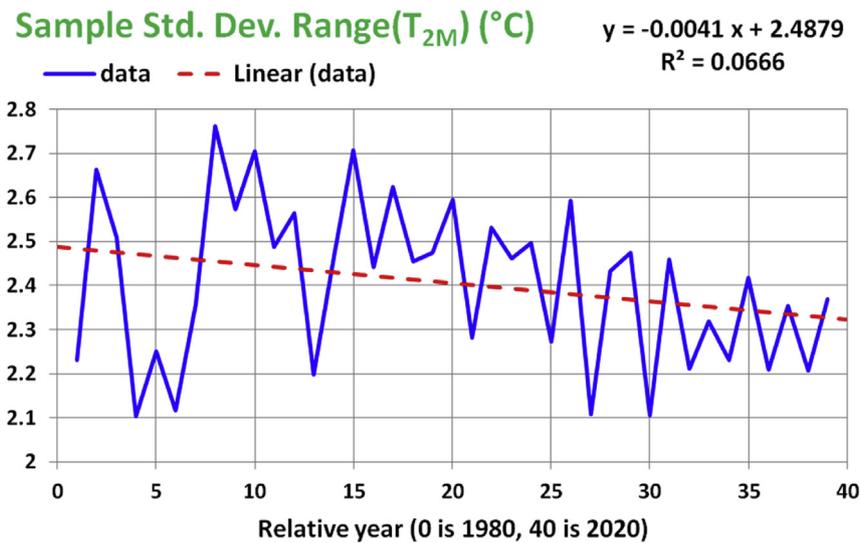

**Figure 7.** Historical data and linear regression line for the annual sample-standard-deviation values of the range of the 2-meter air temperature. The x-axis represents years from 1980 to 2020.

The coefficient of determination (R-squared) as a measure of the usefulness of the regression model (Harrell, 2015) for this regression equation is 0.506, which is nearly midway between the best value of 1 and the worst value of 0. Figure 3 highlights the 95% confidence interval (CI) and the 95% prediction interval (PI) of that regression model for the annual mean of $T_{2M}$. In general, the 95% confidence interval for a linear regression is the band around the regression line where we are 95% confident that the true best-fit regression line (considering more and more input data) lies within it. The 95% prediction interval for a given x value is the band around the predicted y value where we are 95% confident that a true y data value at that x value lies within it (Zaiontz, 2020).

Visual inspection and quantitative assessment of the linear regression (through the R-squared) support its usefulness, and that there is a linear dependence between the temperature and the time on the long term, providing an evidence of a local air warming phenomenon at a rate of about 0.039 °C/year.

To further check the reliability of the above regression model, an ANOVA (analysis of variance) test was performed as in Table 1 (Heumann and Shalabh, 2016). The term (df) is the degrees of freedom, the term (SS) is the sum of squares, the term (MS) is the mean of squares. The p-value is 3.86e-7, which is much smaller than the arbitrarily-set target significance level of α = 0.05. Thus, the regression model is further supported. The reported p-value in Table 1 is (1 – CFD) for the F-distribution at the obtained F-value of 37.89, where CFD is the cumulative density function for the F-distribution, evaluated here with the first degree of freedom being 1 and the second degree of freedom being 37. The F-value is like a signal-to-noise ratio, so a high F-value like the one obtained here is favorable.

**Table 8.** ANOVA test for the linear regression model for the sample standard deviation of range ($T_{2M}$).

|            | df | SS         | MS        | F       | p-value |
|------------|----|------------|-----------|---------|---------|
| Regression | 1  | 0.08368976 | 0.0836898 | 2.63944 | 0.11273 |
| Residual   | 37 | 1.17317245 | 0.0317074 |         |         |
| Total      | 38 | 1.25686221 |           |         |         |

**Table 9.** t-test of significance for the linear regression model for the sample standard deviation of range ($T_{2M}$).

|       | Coefficients | Standard Error | t Stat   | p-value   | Lower 95% | Upper 95% |
|-------|--------------|----------------|----------|-----------|-----------|-----------|
| $b_0$ | 2.48792      | 0.058141       | 42.7909  | 4.074e-33 | 2.37011   | 2.605724  |
| $b_1$ | -0.004116    | 0.002533       | -1.62464 | 0.11273   | -0.009249 | 0.0010173 |

**Table 10.** ANOVA analysis for range ($T_{2M}$) in 9 groups represented by years 1981, 1985, 1990, 1995, 2000, 2005, 2010, 2015, and 2019.

| Source of Variation | SS       | df   | MS       | F      | p-value   | Critical F |
|---------------------|----------|------|----------|--------|-----------|------------|
| Between Groups      | 404.5598 | 8    | 50.56998 | 8.6707 | 8.899e-12 | 1.9412     |
| Within Groups       | 19100.81 | 3275 | 5.832309 |        |           |            |
| Total               | 19505.37 | 3283 |          |        |           |            |

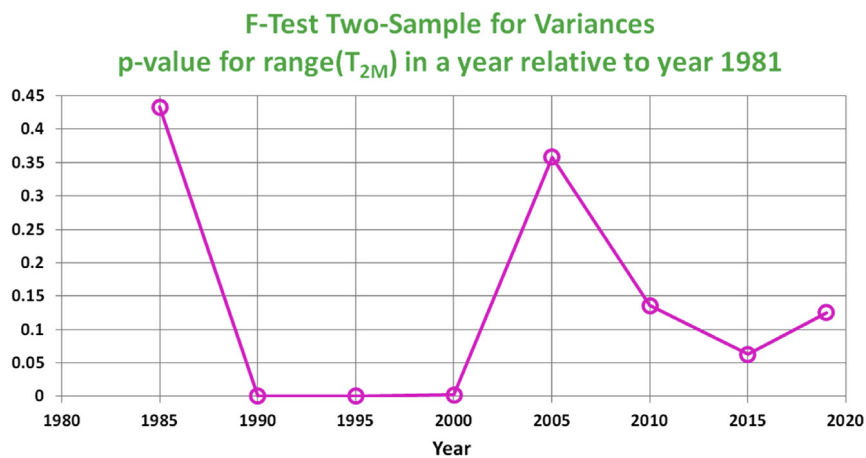

**Figure 8.** p-value of the F-test two-sample for variances of range ($T_{2M}$), comparing the sample variance of each year of: 1985, 1990, 1995, 2000, 2005, 2010, 2015, and 2019 with the sample variance of year 1981.

**Table 11.** t-test for means, two-sample assuming equal variances (pooled version of the t-test for two independent groups), for ($T_{2M}$) in a year relative to year 1981.

| Year Pair | t Statistic | p-value |
|---|---|---|
| 1981, 1985 | 0.0373 | 0.51487 |
| 1981, 1990 | -0.9310 | 0.17608 |
| 1981, 1995 | 0.5395 | 0.70515 |
| 1981, 2000 | -1.2795 | 0.10057 |
| 1981, 2005 | -1.2933 | 0.09815 |
| 1981, 2010 | -2.1218 | 0.01709 |
| 1981, 2015 | -2.2247 | 0.01321 |
| 1981, 2019 | -1.8928 | 0.02939 |

A test of significance for the slope regression coefficient (denoted as $b_1$) and the intercept regression coefficient (denoted as $b_0$), was done through the t-test (Student t-test) as given in Table 2. The p-value for either coefficient is well below 0.05. Also, the band between the lower 95% bound and upper 95% bound excludes the zero value. Both features testify to the statistical significance of both coefficients.

Next, we investigate the sample standard deviation (as a measure of the data spread) of $T_{2M}$ over the same duration of 39 years considered earlier. Figure 4 shows a visual representation of the data points (connected by solid lines), and the linear regression line (as a dashed line). The regression trend line is slightly inclined down on the right end, in agreement with the small negative estimated slope regression coefficient of -0.0032 °C/year. Such value means that the spread in $T_{2M}$ is overall dropping with time, starting from an estimated value of 7.2439 °C in 1980. However, this assumed linear relation is not reliable given the small value of R-squared, being only 0.0168.

The ANOVA table in Table 3 for this linear regression model also does not validate it, given the large p-value of 0.43123, which is well above the significance level of 0.05. Similarly, the t-test of model coefficients in Table 4 shows that the 95% confidence interval for the slope coefficient contains the zero value. So, the proposition that the local 2-meter air temperature becomes less scattered over time is rejected.

The single factor ANOVA analysis for the $T_{2M}$ was performed with time being the factor. Instead of including all 39 years in that analysis, a nominal separation interval of 5 years (reduced to 4 years in the first and last pairs or years) was applied, leading to the selection of 9 years (9 groups) as: 1981, 1985, 1990, 1995, 2000, 2005, 2010, 2015, and 2019. The analysis summary in Table 5 suggests that there is a difference among these groups. Although the p-value (0.04535) is not highly below the significant level ($\alpha = 0.05$), one can still reject the null hypothesis that there is no difference among the groups (no influence of the 4/5 years interval). In the table, the Critical F value is the F-distribution value (with the first degree of freedom being 8 and the second degree of freedom being 3275) at which the area under the tail is equal to the significance level of 0.05. The obtained F is 1.9778, which is above the critical F of 1.9412 because the significance level is above the p-value.

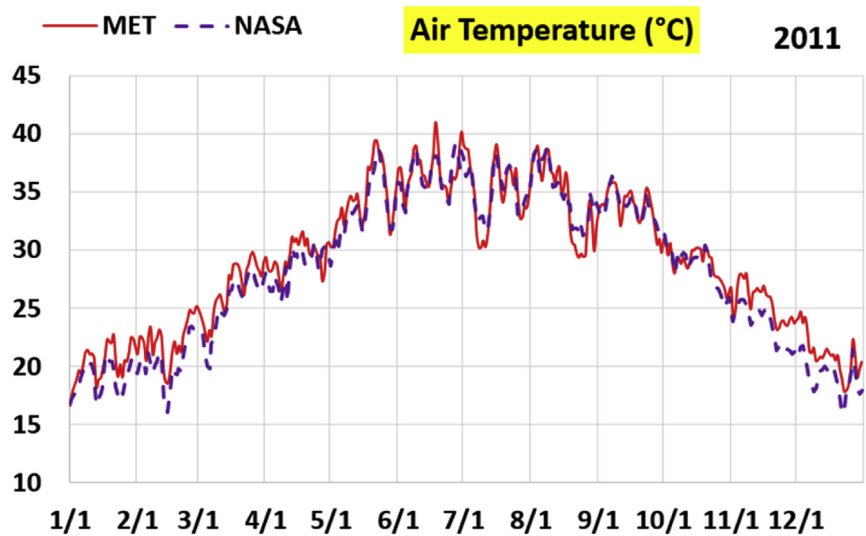

**Figure 9.** Daily air temperatures (at 2 m above ground) from NASA POWER data (NASA) and from measurements at Manah meteorological station (MET) for year 2011.

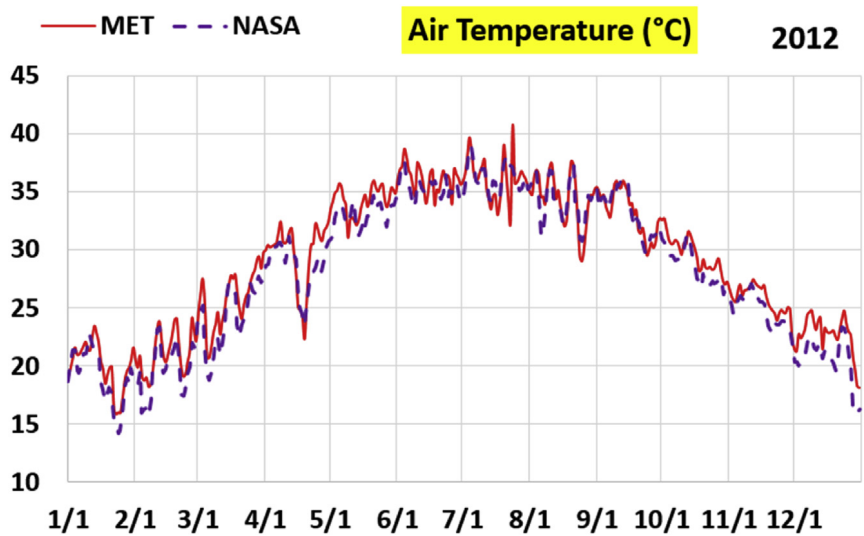

**Figure 10.** Daily air temperatures (at 2 m above ground) from NASA POWER data (NASA) and from measurements at Manah meteorological station (MET) for year 2012.

The above ANOVA table does not refer to a specific dissimilarity between pairs of the 9 years considered or a time influence. A pairwise statistical investigation of the sample variance is done using the F-test for 8 pairs of the above 9 years, by contrasting each of the years except the first one (1981) with the first one. Such approach may help revealing a temporal pattern. The p-values of the 8 F-tests are displayed in Figure 5. None of them falls below the significant level of 0.05 (the lowest is 0.055475, corresponding to the pair of 1981 and 2019). Although the p-value for the last 4 tests show monotonic decline (thus more and more inclination toward an indication of different variance from that of the year 1981), the null hypothesis of equal sample variances cannot be rejected.

Moving to the range of $T_{2M}$, it is noted that when the range is calculated by subtracting the reported minimum of $T_{2M}$ (as a separate variable) from the reported maximum of $T_{2M}$ (as a separate variable), the results are always compatible with the directly reported range of $T_{2M}$ (as a separate variable) within a tolerance of $\pm 0.01$ °C. More specifically, out of the 14,242 data points (also days) considered here; 10,694 (75.1% of data points) showed no mismatch; 1,772 (12.4% of data points) showed a mismatch of -0.01 °C (the calculated range is smaller than the directly reported one), and 1,776 (12.5% of data points) showed a mismatch of 0.01 °C (the calculated result is larger than the directly reported one). With the majority of points showing no mismatch, and with the individual mismatch magnitude is only 0.01 °C, this matter is not considered important. The analysis here is based on processing the directly reported range of $T_{2M}$. A justification of this $\pm 0.01$ °C mismatch can be made in light of the observation that the reported minimum and maximum values pf $T_{2M}$ appear to be recorded with 2 decimal places. With possible rounding, a deviation of $\pm 0.01$ °C becomes expected.

The variation of annual mean for range ($T_{2M}$) over years is presented in Figure 6, along with the linear regression line, regression equation, and the R-squared measure of the quality of the regression model. While the regression trend line and the slope coefficient suggest a long-term decline over years, at an average rate of 0.0067 °C/year, the low R-squared value of 0.0502 does not support the reliability of this model. This is in agreement with the relatively high p-value of 0.17019 in the ANOVA test results in Table 6, and the inclusion of the zero within the 95% confidence interval for the slope coefficient ($b_1$) in Table 7.

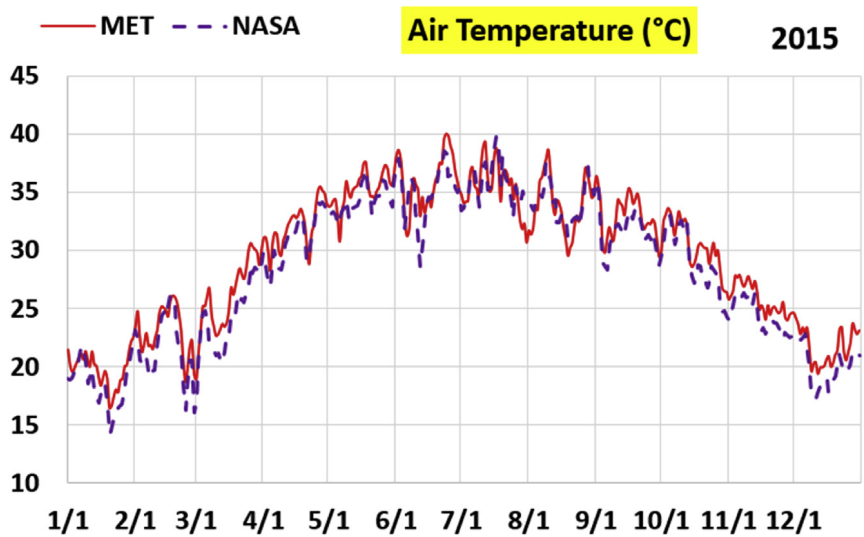

**Figure 11.** Daily air temperatures (at 2 m above ground) from NASA POWER data (NASA) and from measurements at Manah meteorological station (MET) for year 2015.

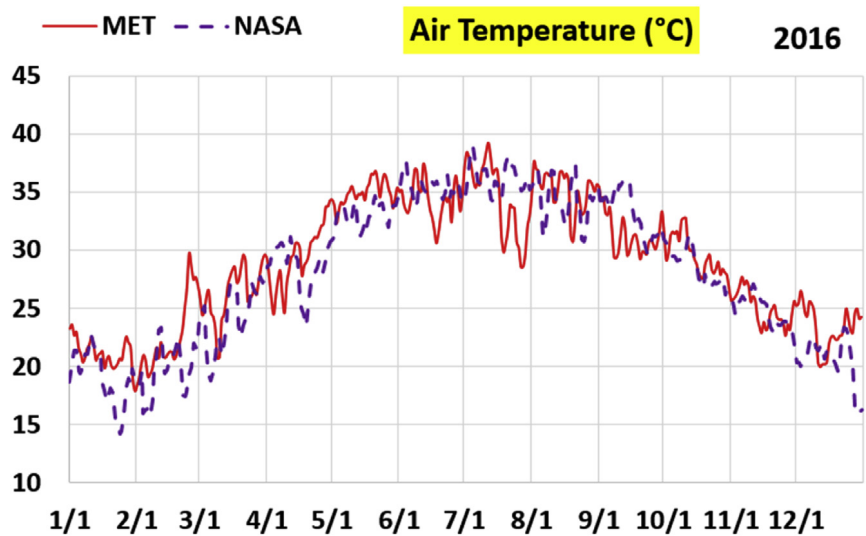

**Figure 12.** Daily air temperatures (at 2 m above ground) from NASA POWER data (NASA) and from measurements at Manah meteorological station (MET) for year 2016.

The linear regression analysis of the standard deviation of the range of $T_{2M}$ is presented in Figure 7, Tables 8 and 9. Similar to the case of the mean of range ($T_{2M}$), the apparent long-term decline rate of 0.0041 °C/year is not statistically significant (low R-squared of 0.0066, high p-value of 0.11273, and the 95% confidence interval for the slope coefficient $b_1$ contains the zero). We cannot reject the null hypothesis (which is that the standard deviation of the range of $T_{2M}$ is independent of time) for the slope coefficient.

The single factor (also called one way) ANOVA analysis for the range of $T_{2M}$, was applied taking 9 years as 9 groups: 1981, 1985, 1990, 1995, 2000, 2005, 2010, 2015, and 2019. The analysis summary is given in Table 10, and it shows that there appears to be a difference among these groups with a very low p-value (8.899e-12) and a relatively high F value of 8.6707. Thus, while there is no support for a linear relation between the range of $T_{2M}$ and time (in years), there is still an indication of influence that is not simply a noise effect.

The F-test for comparing the sample variance of range ($T_{2M}$) in each of the years considered in the ANOVA test except the first year with the sample variance of the first year (1981) was performed. The p-values are presented in Figure 8. The p-values in 3 successive pairs (1981 with 1990, 1995, and 2000) were below 0.002 (which is much smaller that the significance level of 0.05). However, there is no clear monotonic behavior in the figure.

The final stage of analysis is the statistical comparison between the mean for a given year and the first year in the data set (1981), with a separation of 5 years (reduced to 4 in the first and the last comparisons). Thus, the years compared are: 1985, 1990, 1995, 2000, 2005, 2010, 2015, and 2019. These are the same 9 years and same 8 pairs considered earlier in the F-test for comparing the sample variances in Figure 5. Given the overall inability to reject the hypothesis of dissimilar sample variances before, and that the data do not correspond to the same objects, the proper t-test type here is the pooled (or 2 independent groups, with equal variances). One-tail (as compared to two-tail) p-values are reported here because we are interested in the alternative hypothesis that the mean of the second year in each pair is smaller than the mean of the first year (fixed as year 1981), in agreement with an assumption that the mean of $T_{2M}$ decreases over time. The t statistic and the p-value for each pair of years are listed in Table 11. When comparing the mean of $T_{2M}$ year in 1981 with mean of $T_{2M}$ in year 1985 and in year 1995, the t statistic was found to be actually negative, thus the p-value exceeds 0.5 (since the observation was against the assumption that the mean decreases over time). The p-value has been below the target significance level of $\alpha = 0.05$ for the last 3 pairs in the table (thus, when considering years 2010, 2015, and 2015 with respect to year 1981).

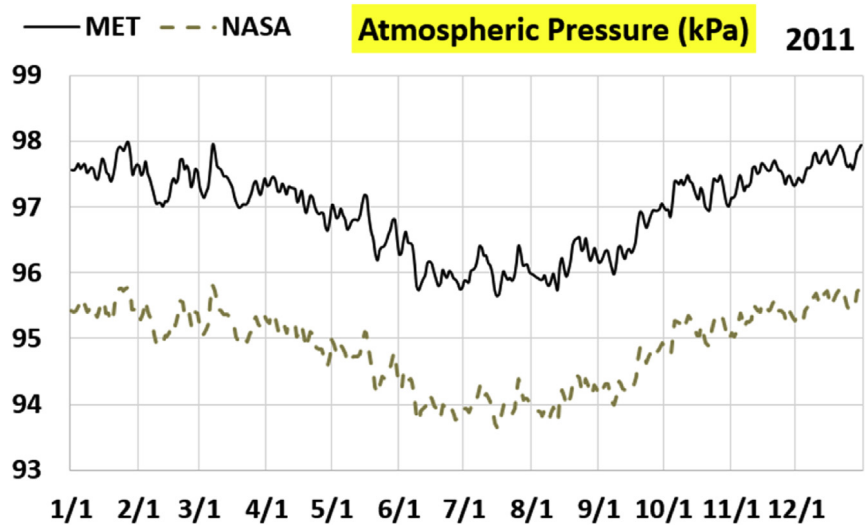

**Figure 13.** Daily atmospheric pressure from NASA POWER data (NASA) and from measurements at Manah meteorological station (MET) for year 2011.

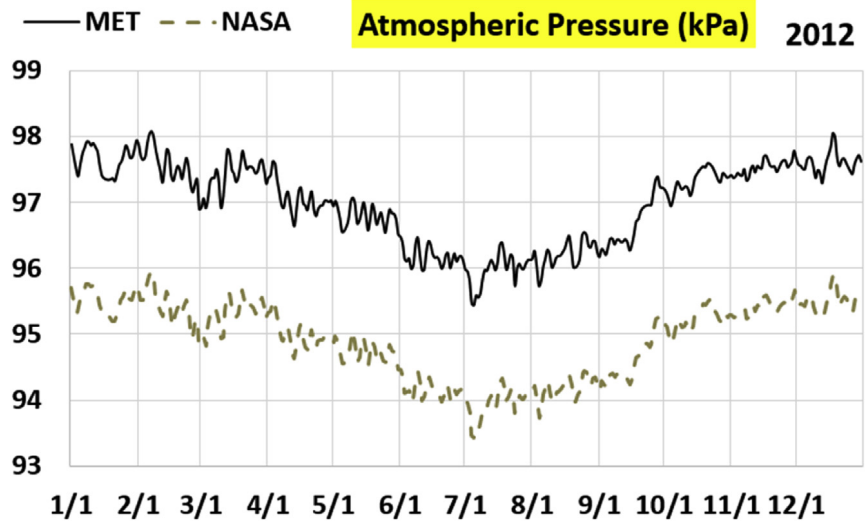

**Figure 14.** Daily atmospheric pressure from NASA POWER data (NASA) and from measurements at Manah meteorological station (MET) for year 2012.

A similar pooled one-tail t-test is not performed for the range of $T_{2M}$, given the linear regression of the mean for that variable that already was not suggesting reliable time dependence.

### 4. Testing NASA POWER ability

In the above analysis of change in local air temperature, it was assumed that the source of data (NASA POWER) is accurate enough to make valuable interpretation. While the author of the database supports its validity, the present section of this work serves as an independent validation of this comprehensive and useful database, by comparing the historical data for 2-meter air temperature with those obtained by actual sensor readings at a specific meteorological station in the Sultanate of Oman. Not only the air temperature at 2 m above ground is considered, but also the atmospheric pressure (station pressure or surface pressure), the relative humidity at 2 m above ground, and the precipitation per day are included in the validation.

The meteorological data collection was managed by the Omani Public Authority of Electricity and Water (PAEW), which later became Public Authority for Water (Diam, 2021) or (Diam) in December 2018. PAEW contracted the Meteorological Office Oman (MOM) to supervise 2 meteorological stations in an effort by the Omani government to assess the feasibility of solar power generation (OPWP, 2012). The two meteorological stations are located in the Omani Government Al Dakhiliyah, with one in the wilayat of Adam, while the other is in the wilayat of Manah (Majlis A'Shura, 2021). We select the meteorological station of Manah to adopt its records as reference true values to compare with. The reason is that as of February 2021, we were able to identify that station using Google satellite maps. The other station of Adam was not located after visual inspection of the satellite map. The longitude and latitude coordinates of the Manah meteorological station as reported by Google

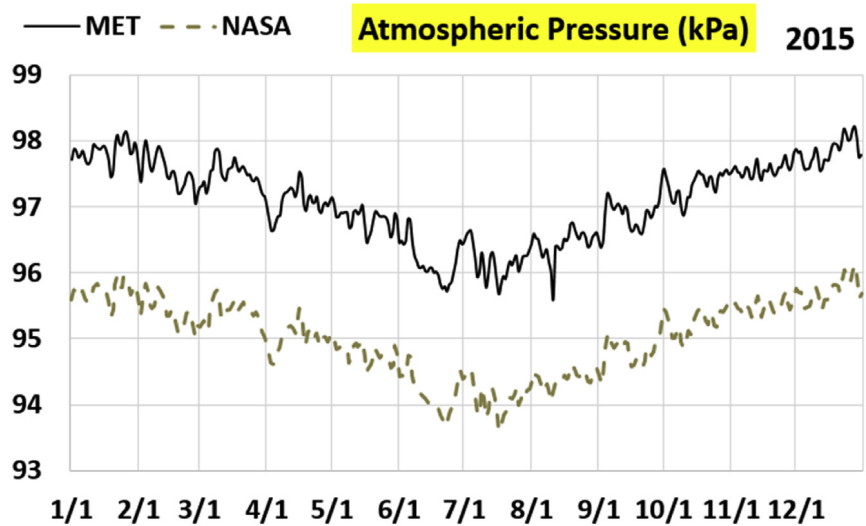

**Figure 15.** Daily atmospheric pressure from NASA POWER data (NASA) and from measurements at Manah meteorological station (MET) for year 2015.

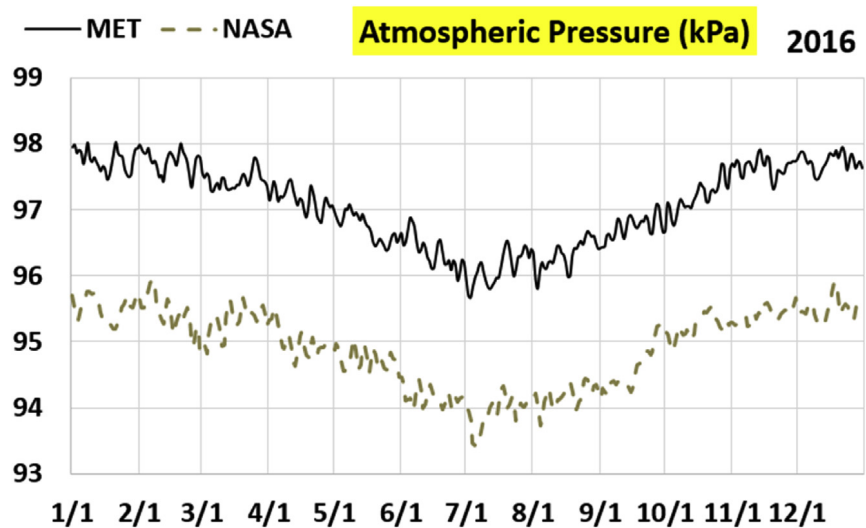

**Figure 16.** Daily atmospheric pressure from NASA POWER data (NASA) and from measurements at Manah meteorological station (MET) for year 2016.

Maps (Google Maps, 2021) are: 22.603121 (North), 57.667213 (East). The sensors used in the station include a HUMICAP temperature and humidity probe (supplier: Vaisala, model: HMP155), a BAROCAP digital barometer (supplier: Vaisala, model: PTB330), and a rain gauge (supplier: Vaisala, model: RG13). An automatic weather station (AWS) (supplier: MicroStep - MIS, model: IMS AMS 111) was used to connect sensors as well as communication equipment of the meteorological station. Data are available freely as one Microsoft Excel file for each year, with two worksheets in each file (one for the Manah station and another for the Adam station). As of 23/February/2021, data exist for years 2011, 2012, 2013, 2104, 2015, and 2016 (OPWP, 2021).

Out of the available six years of data measurements, we excluded the years 2013 and 2014, because a small part of the needed data was missing or showing strange values. So, our validation is based on complete sets of readings for years 2011 (365 days), 2012 (366 days), 2015 (365 days), and 2016 (366 days). The raw data are presented in a tabular form with one row for one hour. Twenty-four records are provided for each day, with equal intervals of one hour. As a post-processing step, the arithmetic mean of these 24 readings was calculated and assigned to the respective day. This converts the hourly records to daily records, which makes them compatible with the NASA POWER data. Another post-processing step was performed, where the pressure of the meteorological station was expressed in kilopascals (kPa) instead of the original unit of hectopascals (hPa), where 1 kPa = 10 hPa. This also makes the data compatible with the NASA POWER outputs, which use the kPa unit for the atmospheric pressure.

The first part of the validation work is conducted by visual inspection of historical profiles of each of the four daily meteorological properties over time. The results are presented in Figures 9, 10, 11, and 12 for the air temperature at 2 m above the ground in degrees Celsius (°C), in Figures 13, 14, 15, and 16 for the atmospheric pressure in kilopascals (kPa), in Figures 17, 18, 19, and 20 for the relative humidity (the percentage of the existing water vapor pressure to the maximum water vapor pressure possible without condensation and dew) at 2 m above the ground, and in Figures 21, 22, 23, and 24 for the precipitation in millimeters per day (mm or mm/day).

The air temperature curves show the best agreement among the two sources of data, not just at the level of average changes over the whole year, but also at the level of fast changes at the level of days (for example, the sharp increase in temperature on 18/June/2011 is identified in both data sources). This is a favorable result, as it supports the correctness of the conducted warming assessment part of this work, and that the adopted NASA POWER data there are reliable.

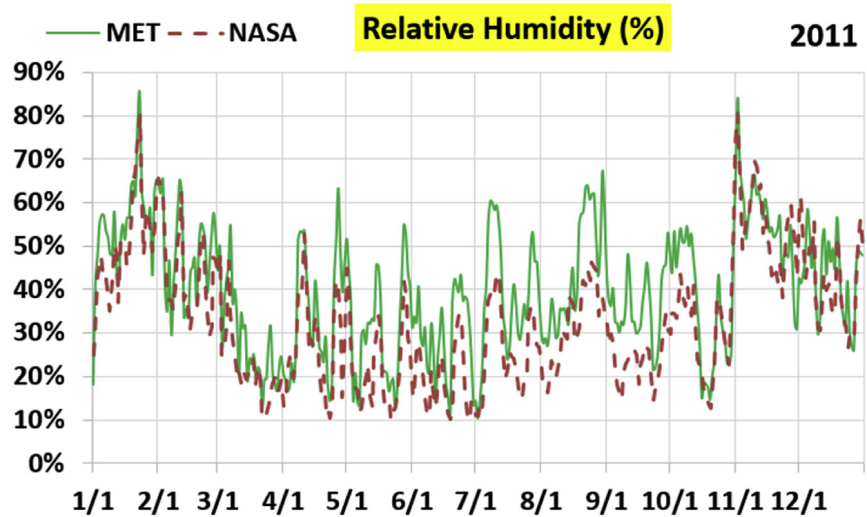

**Figure 17.** Daily relative humidity (at 2 m above ground) from NASA POWER data (NASA) and from measurements at Manah meteorological station (MET) for year 2011.

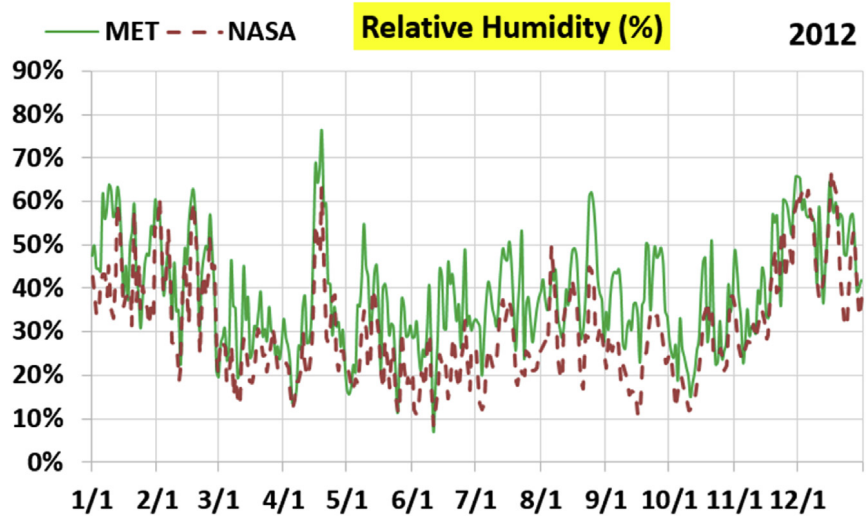

**Figure 18.** Daily relative humidity (at 2 m above ground) from NASA POWER data (NASA) and from measurements at Manah meteorological station (MET) for year 2012.

The atmospheric pressure curves suggest that NASA POWER data successfully capture the temporal variation, but has a persistent offset of about 2 kPa. The NASA POWER pressures is below the recorded ones for all 4 years studied. One may think that this offset is because of recording barometric pressure (equivalent pressure at sea level) as compared to actual pressure at the station. However, this explanation is rejected because the NASA POWER pressures are below the measured ones, whereas they should be above the measured ones in case of being barometric rather than atmospheric (Struchtrup, 2014). It is worth mentioning that Manah meteorological station is at an elevation of about 344 m (Daft Logic, 2021). This corresponds to a pressure drop of 4.064 kPa below the sea level value, assuming standard atmospheric variation with elevation (NOAA-NASA-USAF, 1976; digital dutch, 2021).

The relative humidity curves do not suffer from a constant offset as was the case with the pressure, but unlike the temperature, there is no good overall agreement between the NASA POWER data points and the measurement-based ones. Both curves vary over a similar extent, but the variation patterns are in mild disagreement.

The daily precipitation (which is practically rain, as snow is not typically observed in Oman) shows the largest deviation between the NASA POWER data and the meteorological sensors data. NASA POWER values are severely above the reality, with the NASA POWER curves obscuring the sensors data curves. The performance of NASA POWER for precipitation in this validation study is not acceptable.

The qualitative comparisons between NASA POWER data and meteorological sensors data for the four weather variables: air temperature at 2 m, atmospheric pressure, relative humidity at 2 m, and daily precipitation presented earlier are supported here by quantitative evaluation of the level of deviation between the two data sets, with the sensors data taken as the true reference one.

Tables 12, 13, 14, and 15 compare 4 statistical values for each of the four climate variables, when calculated from the NASA POWER (data denoted by 'NASA') and when calculated from the meteorological sensors' readings (data denoted by 'MET'). These statistical values are the maximum, the minimum, the arithmetic mean (simple average), and the population standard deviation (a measure of scatter in the data and their deviation from their mean). One table is provided per year. The air temperature shows small numerical differences for all four statistical values, making NASA POWER data reliable for this particular variable.

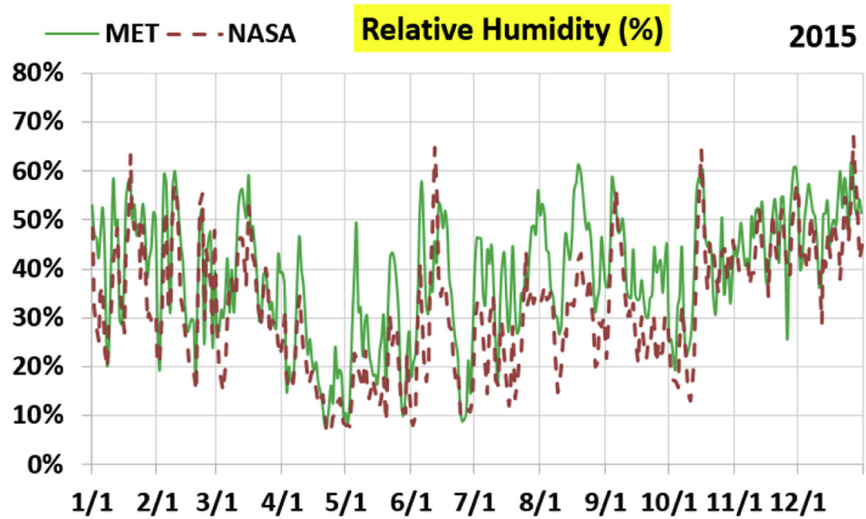

**Figure 19.** Daily relative humidity (at 2 m above ground) from NASA POWER data (NASA) and from measurements at Manah meteorological station (MET) for year 2015.

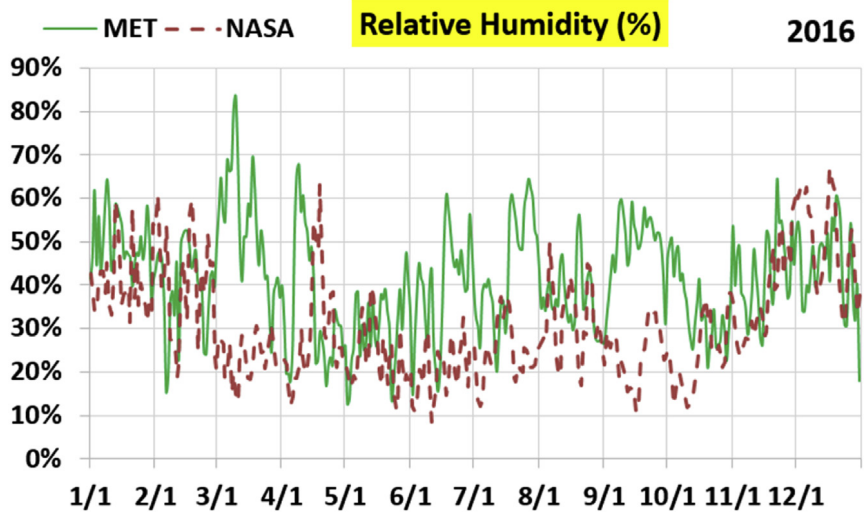

**Figure 20.** Daily relative humidity (at 2 m above ground) from NASA POWER data (NASA) and from measurements at Manah meteorological station (MET) for year 2016.

For the atmospheric pressure, a near 2.1 kPa difference exists between the two sets for all 4 years. However, the range (the maximum value minus the minimum value) and the standard deviation are similar for both data sets. The relative humidity shows similar standard deviations from the NASA POWER data and from the sensors data, but the differences in the mean, maximum, and minimum are not negligible. The serious error in the precipitation as obtained from NASA POWER data is clearly manifested in these tables. For example, the 2011 mean precipitation according to NASA POWER was 0.404 mm (rounded in the table to 0.40), which is 15.9 times the measured value of only 0.02546 mm (rounded in the table to 0.03). The ratio of precipitation means in 2012 is 38.0, and for 2015 is 90.2, and for 2016 is 19.4. With no exception, each of the four years considered shows failure in obtaining realistic intensity of rain. The true situation is much drier than the one suggested by NASA POWER.

The quantitative evaluation of mismatch between NASA POWER data and the meteorological sensors data for the four weather variables continue in Tables 16, 17, 18, and 19. For each year, a table shows two numbers for each variable. One number is the arithmetic mean of the absolute difference for the daily weather variable between the values based on NASA POWER data and the values based on the meteorological station data. The other number is the root-mean-square (RMS) of these daily records of differences, which is the square root of the average squared daily differences. Both numbers measure the disparity between the two sets of data. They become equal in the special case when the differences are identical in all days, this is nearly the case in the pressure, and this agrees with the visually observed near-constant shift earlier when the data were represented as curves. On the other hand, the daily absolute differences in the precipitation variable show a much larger RMS than the mean, which indicates irregularity of these daily differences. For the air temperature, the two numbers expressing disparity levels are within 3.1 °C (within 1.7 °C if excluding year 2016). This is not a huge gap. On the other hand, the disparity in the relative humidity is within 20.1% (11.2% if excluding year 2016).

## 5. Justification of parametric statistical methods

In statistical analysis, one may encounter two families of test methods: parametric test methods and non-parametric test methods. In parametric tests, a certain probability distribution (especially the normal distribution) of data is implicitly assumed as a prerequisite for high results validity. On the other hand, non-parametric methods (or distribution-free methods) do not require a certain distributional condition. In the present work, a mix of parametric tests was used, including the two-sample t-test for means, and the ANOVA test for means, and both of them assume normal distribution of data (Hoskin, 2021).

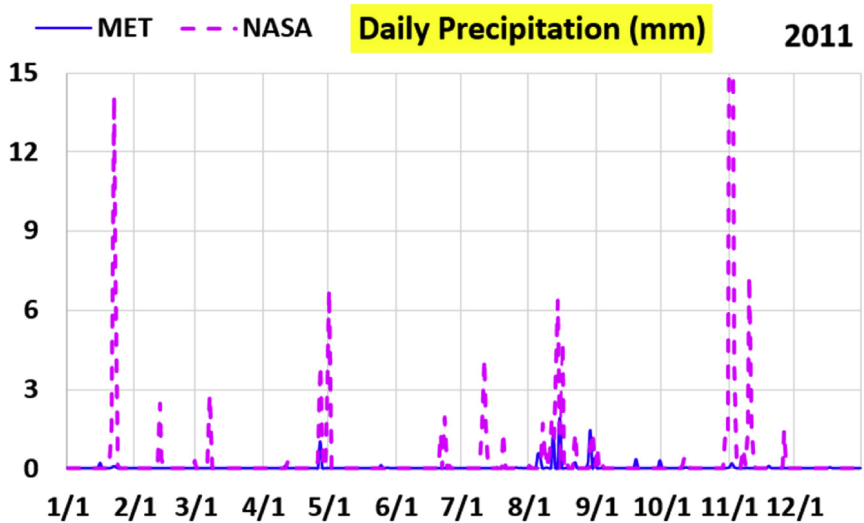

**Figure 21.** Daily precipitation from NASA POWER data (NASA) and from measurements at Manah meteorological station (MET) for year 2011.

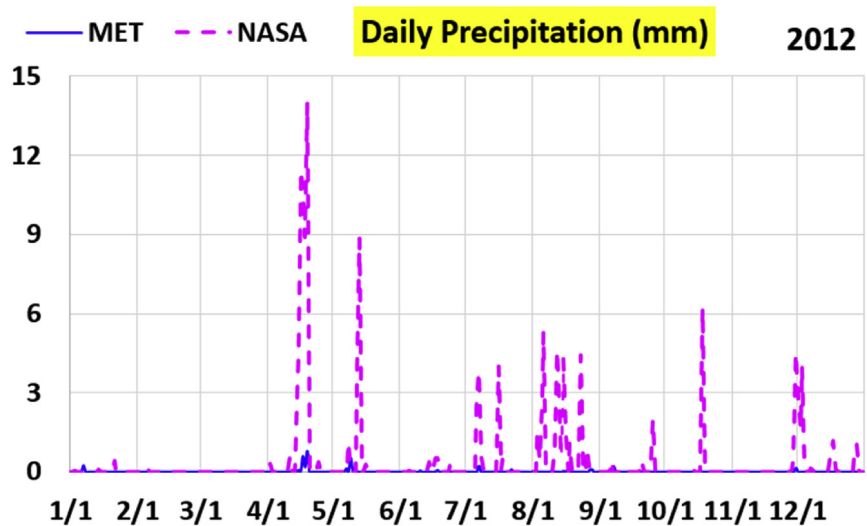

**Figure 22.** Daily precipitation from NASA POWER data (NASA) and from measurements at Manah meteorological station (MET) for year 2012.

While no normality test was presented so far, such action is not necessary here given the relatively large size of samples (being either 365 or 366 daily data points per group). This is safely large, while attention to satisfying the normality condition is applicable only for small samples of about 20 points of less (Winthrop University Hospital, 2021; van den Berg, 2021). The F-test also assumes data normality. However, the test is robust for equal sample sizes with a suitable size, such as 32 (Donaldson, 1966).

For the identified linear regression model about the local annual mean air temperature at 2 m above ground, reliable use of the ANOVA and t-test analyses of the regression coefficients assume that the predication residuals (errors) are normally distributed with a mean of zero but do not require that this temperature by itself is normally distributed. Moreover, some deviation from normality by the predication residuals errors can be tolerated for reasonable sample sizes, such as 30 points and more (Kim, 2015). The normality assessment of the residuals (a residual value is defined as: the observed temperature minus the predicted temperature from the regression model at a given year) was performed quantitatively and qualitatively, and the results are presented in this section.

Table 20 lists all 39 residual values. They range from -0.99483 °C to 1.05559 °C. Their mean is 5.10133e-15 °C, which is practically zero, and satisfies the zero-mean condition.

The first inspection of normality of the regression residuals is quantitative, and it is expressed by the closeness of the median to the mean. The median was found to be 0.06168 °C. The range (maximum – minimum) of the residuals is 2.05042 °C, so the distance between the mean and median is only 3.01% of the range. Thus, the first check of the normality of the regression residuals of the local annual-mean values of the 2-meter air temperature led to a favorable outcome.

The scatter plot of the residual errors versus the predicted values of the local annual-mean air temperature at 2 m above ground ($T_{2M}$) is given in Figure 25. There is no obvious pattern, and the errors show some randomness in their sign and magnitude. This is the second way to inspect the normality of the regression residuals, but in a qualitative way. The outcome is also favorable.

The third inspection of normality of the regression residuals is qualitative, and it is expressed through comparing the histogram (in terms of

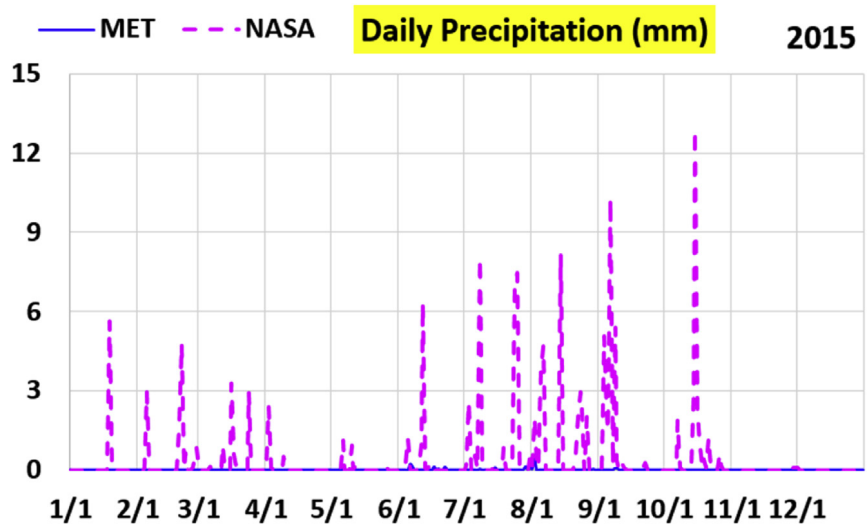

**Figure 23.** Daily precipitation from NASA POWER data (NASA) and from measurements at Manah meteorological station (MET) for year 2015.

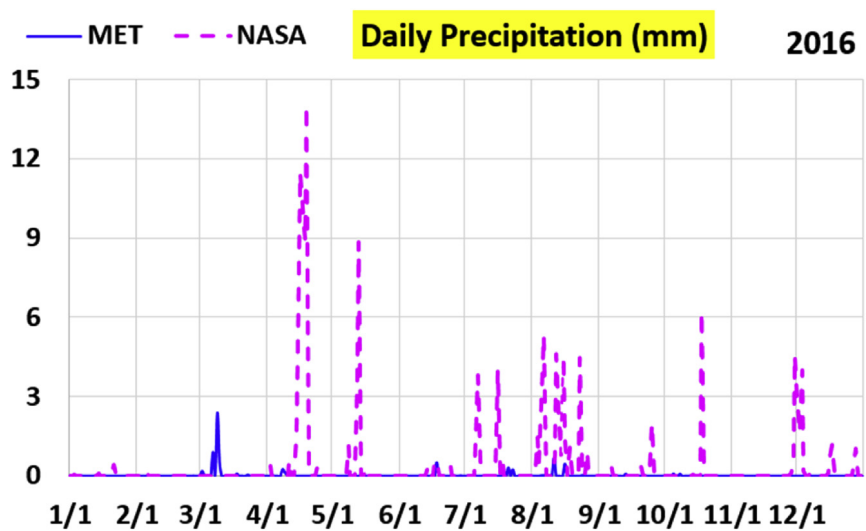

**Figure 24.** Daily precipitation from NASA POWER data (NASA) and from measurements at Manah meteorological station (MET) for year 2016.

**Table 12.** Comparison of statistical data for four weather properties from meteorological sensors recordings (MET) and from NASA POWER (NASA) at Manah meteorological station, by analyzing daily data for 2011.

| Property | Source | Maximum | Minimum | Arithmetic mean | Population standard deviation |
|---|---|---|---|---|---|
| Air temperature, at 2 m (°C) | MET | 40.94 | 16.60 | 29.02 | 6.00 |
|  | NASA | 39.07 | 16.00 | 28.21 | 6.58 |
| Atmospheric pressure (kPa) | MET | 97.98 | 95.64 | 96.93 | 0.64 |
|  | NASA | 95.81 | 93.64 | 94.86 | 0.60 |
| Relative humidity, 2 m (%) | MET | 85.29% | 10.46% | 40.12% | 14.48% |
|  | NASA | 80.54% | 9.87% | 32.81% | 14.88% |
| Precipitation (mm/day) | MET | 1.94 | 0.00 | 0.03 | 0.16 |
|  | NASA | 28.33 | 0.00 | 0.40 | 2.13 |

**Table 13.** Comparison of statistical data for four weather properties from meteorological sensors recordings (MET) and from NASA POWER (NASA) at Manah meteorological station, by analyzing daily data for 2012.

| Property | Source | Maximum | Minimum | Arithmetic mean | Population standard deviation |
|---|---|---|---|---|---|
| Air temperature, at 2 m (°C) | MET | 40.78 | 15.88 | 29.14 | 5.95 |
|  | NASA | 38.92 | 14.17 | 28.16 | 6.35 |
| Atmospheric pressure (kPa) | MET | 98.08 | 95.45 | 97.00 | 0.63 |
|  | NASA | 95.95 | 93.42 | 94.91 | 0.59 |
| Relative humidity, 2 m (%) | MET | 76.42% | 7.00% | 39.00% | 12.39% |
|  | NASA | 66.97% | 8.51% | 30.67% | 12.36% |
| Precipitation (mm/day) | MET | 0.78 | 0.00 | 0.01 | 0.06 |
|  | NASA | 14.00 | 0.00 | 0.38 | 1.49 |

**Table 14.** Comparison of statistical data for four weather properties from meteorological sensors recordings (MET) and from NASA POWER (NASA) at Manah meteorological station, by analyzing daily data for 2015.

| Property | Source | Maximum | Minimum | Arithmetic mean | Population standard deviation |
|---|---|---|---|---|---|
| Air temperature, at 2 m (°C) | MET | 40.01 | 16.44 | 29.41 | 5.90 |
|  | NASA | 39.77 | 14.21 | 28.31 | 6.35 |
| Atmospheric pressure (kPa) | MET | 98.23 | 95.60 | 97.10 | 0.62 |
|  | NASA | 96.12 | 93.59 | 95.02 | 0.58 |
| Relative humidity, 2 m (%) | MET | 61.66% | 7.92% | 38.73% | 12.83% |
|  | NASA | 66.95% | 6.34% | 31.62% | 13.11% |
| Precipitation (mm/day) | MET | 0.42 | 0.00 | 0.00 | 0.03 |
|  | NASA | 12.87 | 0.00 | 0.40 | 1.42 |

**Table 15.** Comparison of statistical data for four weather properties from meteorological sensors recordings (MET) and from NASA POWER (NASA) at Manah meteorological station, by analyzing daily data for 2016.

| Property | Source | Maximum | Minimum | Arithmetic mean | Population standard deviation |
| --- | --- | --- | --- | --- | --- |
| Air temperature, at 2 m (°C) | MET | 39.28 | 17.84 | 28.98 | 5.41 |
|  | NASA | 38.92 | 14.17 | 28.16 | 6.35 |
| Atmospheric pressure (kPa) | MET | 98.03 | 95.66 | 97.07 | 0.62 |
|  | NASA | 95.95 | 93.42 | 94.91 | 0.59 |
| Relative humidity, at 2 m (%) | MET | 83.41% | 12.56% | 41.49% | 12.29% |
|  | NASA | 66.97% | 8.51% | 30.67% | 12.36% |
| Precipitation (mm/day) | MET | 2.37 | 0.00 | 0.02 | 0.15 |
|  | NASA | 14.00 | 0.00 | 0.38 | 1.49 |

**Table 16.** Numerical level of absolute daily deviation for four weather properties between meteorological sensors recordings (MET) and NASA POWER (NASA) at Manah meteorological station, for year 2011.

| Property | Arithmetic mean of \|NASA – MET\| array | Root mean square of \|NASA – MET\| array |
| --- | --- | --- |
| Air temperature, at 2 m (°C) | 1.23 | 1.47 |
| Atmospheric pressure (kPa) | 2.07 | 2.07 |
| Relative humidity, 2 m (%) | 8.78% | 10.71% |
| Precipitation (mm/day) | 0.41 | 2.16 |

**Table 17.** Numerical level of absolute daily deviation for four weather properties between meteorological sensors recordings (MET) and NASA POWER (NASA) at Manah meteorological station, for year 2012.

| Property | Arithmetic mean of \|NASA – MET\| array | Root mean square of \|NASA – MET\| array |
| --- | --- | --- |
| Air temperature, at 2 m (°C) | 1.27 | 1.50 |
| Atmospheric pressure (kPa) | 2.08 | 2.08 |
| Relative humidity, 2 m (%) | 9.20% | 11.18% |
| Precipitation (mm/day) | 0.37 | 1.50 |

relative frequency) of the residuals with a scaled normal distribution curve whose mean is the mean of the residual errors (which is nearly zero), and whose standard deviation is equal to the sample standard deviation of the residual errors (which is 0.43674 °C). The histogram is given in Figure 26. This histogram has 11 bins of equal widths of 0.2 °C. The continuous curve of normal distribution in the figure is scaled vertically such that its peak matches the peak of the histogram, and this is achieved by multiplying the raw normal probability curve by a factor of 0.25263. While the histogram is left-skewed and does not perfectly follow the scaled normal distribution curve, it also does not deviate significantly from it. It should be noted that the data size for the regression is 39, which is not small and such mild deviation from perfect normality is not considered a source of concern.

The last inspection tool of normality for regression residuals of the local annual-mean values of the 2-meter air temperature ($T_{2M}$) is also qualitative, and is performed in Figure 27 by visually contrasting the perfect straight line to the curve representing the actual residuals but projected onto the Z statistic of the standard normal distribution. In forming this projection, a cumulative distribution factor was constructed, which neither takes exactly the value of 0 nor takes exactly the value of 1. However, it starts at (1/(2N)) = 0.01282, where N is the number of residual points (39), and ends at (1 - 1/(2N)) = 0.98718. Between these extreme points, the cumulative distribution factor is incremented with equal spaces of $(1/N) = 0.02564$. Then, the values of the horizontal axis are the Z scores where the cumulative distribution function values of the standard normal distribution correspond to the given cumulative distribution factor values. The values of the vertical axis for the straight line are obtained by finding the non-standard (using the mean and sample standard deviation of the regression residuals) normal distribution at each corresponding cumulative distribution factor. The residual values are ordered ascendingly before plotting then versus the Z scores (thus, both have non-decreasing profiles).

**Table 18.** Numerical level of absolute daily deviation for four weather properties between meteorological sensors recordings (MET) and NASA POWER (NASA) at Manah meteorological station, for year 2015.

| Property | Arithmetic mean of \|NASA – MET\| array | Root mean square of \|NASA – MET\| array |
| --- | --- | --- |
| Air temperature, at 2 m (°C) | 1.39 | 1.62 |
| Atmospheric pressure (kPa) | 2.08 | 2.08 |
| Relative humidity, 2 m (%) | 8.35% | 10.17% |
| Precipitation (mm/day) | 0.40 | 1.48 |

**Table 19.** Numerical level of absolute daily deviation for four weather properties between meteorological sensors recordings (MET) and NASA POWER (NASA) at Manah meteorological station, for year 2016.

| Property | Arithmetic mean of \|NASA – MET\| array | Root mean square of \|NASA – MET\| array |
| --- | --- | --- |
| Air temperature, at 2 m (°C) | 2.35 | 3.05 |
| Atmospheric pressure (kPa) | 2.16 | 2.17 |
| Relative humidity, at 2 m (%) | 15.96% | 20.07% |
| Precipitation (mm/day) | 0.39 | 1.54 |

Table 20. Residual errors for the identified linear regression model for the local annual-mean values of the 2-meter air temperature.

| Year | Residual (°C) | Year | Residual (°C) | Year | Residual (°C) |
|---|---|---|---|---|---|
| 1981 | 0.33650 | 1994 | 0.10870 | 2007 | -0.00269 |
| 1982 | -0.99483 | 1995 | -0.47867 | 2008 | -0.37028 |
| 1983 | -0.98184 | 1996 | -0.17469 | 2009 | -0.06343 |
| 1984 | 0.06118 | 1997 | -0.43293 | 2010 | 0.29624 |
| 1985 | 0.16255 | 1998 | 1.05559 | 2011 | -0.13296 |
| 1986 | -0.00931 | 1999 | 0.71586 | 2012 | 0.12076 |
| 1987 | 0.43313 | 2000 | 0.26023 | 2013 | -0.61570 |
| 1988 | 0.44571 | 2001 | 0.25461 | 2014 | -0.48057 |
| 1989 | -0.11999 | 2002 | 0.25825 | 2015 | 0.15864 |
| 1990 | 0.47469 | 2003 | 0.21648 | 2016 | -0.49310 |
| 1991 | 0.19563 | 2004 | 0.32481 | 2017 | -0.04062 |
| 1992 | -0.87993 | 2005 | 0.06168 | 2018 | 0.29629 |
| 1993 | 0.14202 | 2006 | 0.03708 | 2019 | -0.14508 |

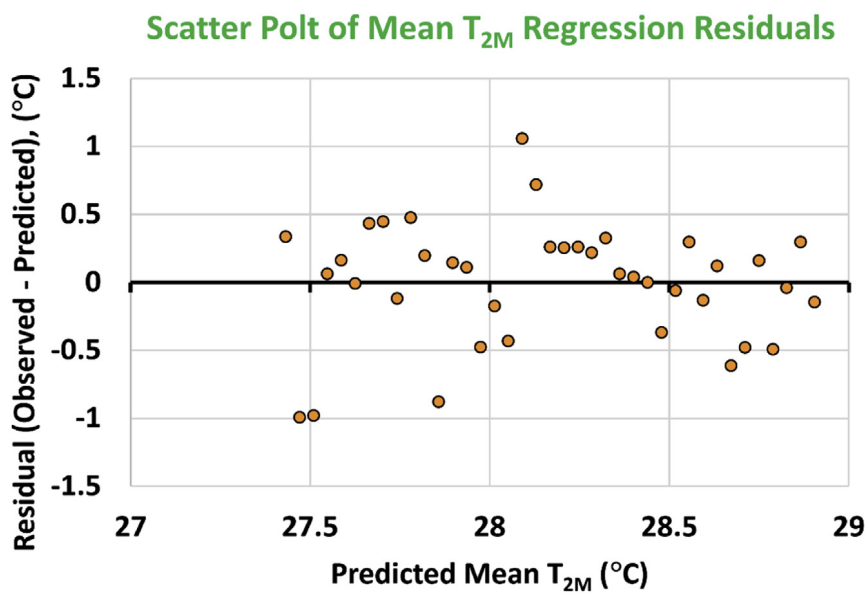

Figure 25. Residual errors for the identified linear regression model for the local annual-mean values of the 2-meter air temperature, plotted versus the predicted values.

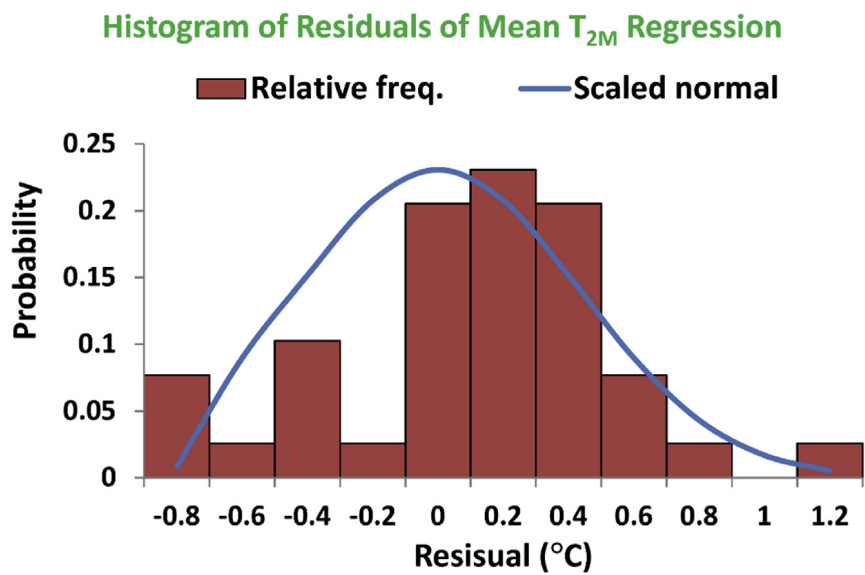

Figure 26. Residual errors for the identified linear regression model for the local annual-mean values of the 2-meter air temperature, plotted as a histogram of relative frequencies with 11 bins. The continuous curve represents a scaled normal distribution profile to compare with.

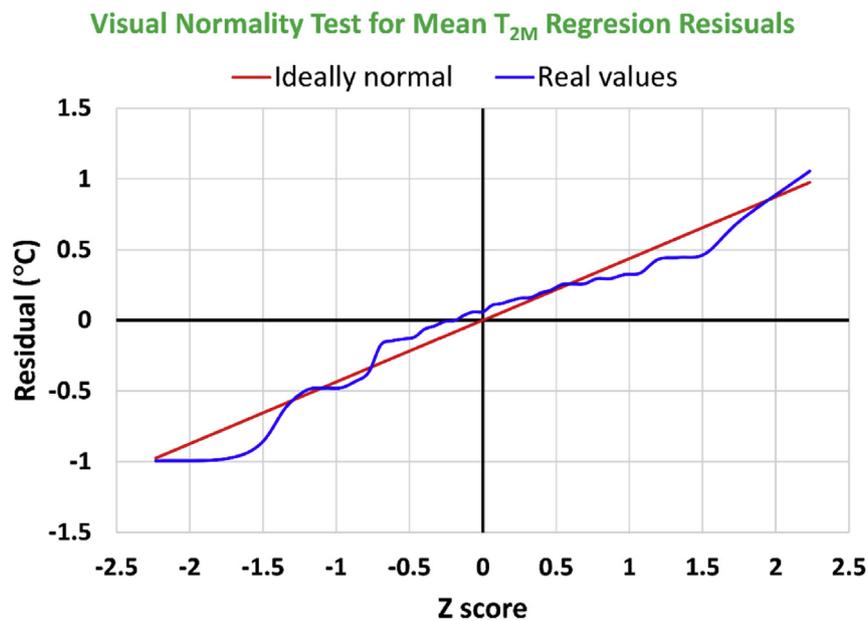

**Figure 27.** Residual errors for the identified linear regression model for the local annual-mean values of the 2-meter air temperature, plotted versus virtual Z scores. The straight line corresponds to perfect normal distribution.

While similar residuals normality inspection can be done for the 39 values corresponding to each of the other three linear regression models presented earlier in this work, the p-value (hypothesis testing) for them did not show statistical significance of time influence, and thus there is no need to support the conformance of the residuals to the normality condition and the zero-mean condition.

## 6. Conclusions

Using a number of statistical tools (linear regression, ANOVA, F-test, and t-test), the freely available contiguous data from NASA (under the POWER project) for air temperature at a height of 2 m above ground in a single point (latitude: 24.233935° N, longitude: 55.892071° E) at Al Buraimi district (wilayah or wilayat) within Al Buraimi Governate in the Sultanate of Oman was investigated. A total of 14,242 data points corresponding to 14,242 days covering the period from 3/January/1981 to 31/December/2019 (inclusive) were considered after dividing them by years. In summary, the following points may be made: the 2-meter air temperature shows a long-term increase at an average rate of approximately 0.039 °C per year, corresponding to about 1.5 °C over 4 decades. The sample standard deviation of the 2-meter air temperature does not show a reliable change over time. There is no enough evidence that the mean and the sample standard deviation of the range of the 2-meter air temperature are time dependent. Comparing the years 1981, 1985, 1990, 1995, 2000, 2005, 2010, 2015, and 2019; they do not seem to be statistically the same in terms of the 2-meter air temperature or its range.

Comparison between daily values of four meteorological properties from NASA POWER and from meteorological station measurements over 4 years suggest that the air temperature data (at 2 m above ground) as obtained from NASA POWER are reliable. Although the other three meteorological properties (atmospheric pressure, relative humidity, and daily precipitation) are outside the scope of the warming assessment part of this work, their inclusion helps in making the work of broader usefulness, where it now has dual purposes: the local assessment of warming (which may attract attention of local readers), and the validation of the NASA POWER data (which may attract global readers, although the validation is done at a single point). Even the part about local assessment of warming may attract global readers since the procedure of analysis is not limited to a specific geographic zone.

### Declarations

*Author contribution statement*

Osama A. Marzouk: Conceived and designed the analysis; Analyzed and interpreted the data; Wrote the paper.

*Funding statement*

This research did not receive any specific grant from funding agencies in the public, commercial, or not-for-profit sectors.

*Data availability statement*

Data associated with this study is available at Prediction Of Worldwide Energy Resources (POWER) by NASA.

*Declaration of interests statement*

The authors declare no conflict of interest.

*Additional information*

No additional information is available for this paper.